\documentclass[]{JHEP3}

\usepackage{epsfig,multicol,bbm}

\newcommand\fverb{\setbox\fverbbox=\hbox\bgroup\verb}
\newcommand\fverbdo{\egroup\medskip\noindent%
            \fbox{\unhbox\fverbbox}\ }
\newcommand\fverbit{\egroup\item[\fbox{\unhbox\fverbbox}]}
\newbox\fverbbox

\newcommand{\eq}[1]{\begin{equation}#1\end{equation}}
\newcommand{\lrp}[1]{\left( #1 \right)}  

\def\be{\begin{equation}}
\def\ee{\end{equation}}
\def\bea{\begin{eqnarray}}
\def\eea{\end{eqnarray}}

\def\bp{{\bf p}}

\def\e{\epsilon}
\def\z{\zeta}
\def\s{\sigma}
\def\a{\alpha}
\def\b{\beta}
\def\y{\eta}
\def\d{\delta}

\def\f{\frac}

\def\ma{\mathcal}
\def\la{\langle}
\def\ra{\rangle}
\def\l{\left}
\def\r{\right}

\title{Primordial Trispectrum from Entropy Perturbations in Multifield DBI Model}

\author{Xian Gao, Bin Hu\\
    Key Laboratory of Frontiers in Theoretical Physics,
    Institute of Theoretical Physics, Chinese Academy of Sciences,
    P.O. Box 2735, Beijing 100190, China\\
    E-mail: \email{gaoxian@itp.ac.cn}, \email{hubin@itp.ac.cn}}

\received{\today}       
\accepted{\today}       

\preprint{CAS-KITPC/ITP-120}

\abstract{We investigate the primordial trispectra of the general
multifield DBI inflationary model. In contrast with the single field
model, the entropic modes can source the curvature perturbations on
the super horizon scales, so we calculate the contributions from the
interaction of four entropic modes mediating one adiabatic mode to
the trispectra, at the large transfer limit ($T_{RS}\gg1$). We
obtained the general form of the 4-point correlation functions,
plotted the shape diagrams in two specific momenta configurations,
``equilateral configuration" and ``specialized configuration". Our
figures showed that we can easily distinguish the two different
momenta configurations.}

\begin{document}


\section{Introduction}
The Cosmic Microwave Background (CMB) provides us with remarkably
detailed signatures of the early universe. The observations from
large scale structures are consistent with an almost scale
invariant, Gaussian primordial density perturbations generated
during inflation. Precision measurements of any small deviation from
Gaussian distribution enables us to distinguish different
cosmological models. In the scalar field(s) inflation case, the
non-Gaussian fluctuations can be parametrized by $f_{NL}$ at the
leading order and $\tau_{NL}$ at sub-leading order respectively. The
current experimental bound for the bispectrum (the three point
correlation function of the primordial curvature perturbation
$\zeta$) is $-9<f^{\textrm{local}}_{NL}<111$ from WMAP5 \cite{fNL},
and for the trispectrum (four point correlation function) is
$|\tau_{NL}|<10^8$ \cite{tauNL1}, the next generation of experiments
such as PLANCK will increase the sensitivity to about
$\tau_{NL}\sim560$ \cite{tauNL2}.

On the theoretical aspect, models with non-Gaussianities have been
intensively investigated in recent years (see \cite{Bartolo:2004if}
for a review). Standard single-field slow-roll inflationary model
predicts an almost Gaussian fluctuation with undetectable
non-Gaussianity \cite{maldacena}. 3-point functions, or its Fourier
transformation the bispectra for single-field and multifiled in
slow-roll models are investigated in \cite{Seery1,Seery2,random}.
Higher-order correlation function, e.g. the trispectra (Fourier
transformation of connected 4-point function) are studied in
\cite{Koyama,Seery3,Seery4,Byrnes:2006vq,Engel:2008fu,Huang:2009xa}.
For k-inflation models \cite{kinf}, DBI inflation models
\cite{Tong,Chen1,Shiu,Tye} and curvaton scenario \cite{Lyth:2001nq},
the primordial bispectra and trispectra are calculated in
\cite{Chen2,Langlois1,Langlois2,Langlois:2009ej,Gao,Huang,running,Koyama2},
in \cite{dbi tri,timedepcs} and in
\cite{Bartolo:2003jx,Huang:2008ze} respectively. And the loop
corrections to the power spectrum and bispectrum
\cite{Seery5,Dimastrogiovanni,Cogollo:2008bi}, the non-Gaussianities
originated in noncommutative effect \cite{noncommu}, $\alpha$ vacuum
\cite{alpha}, thermal fluctuations \cite{thermal}, string gas
\cite{stringgas}, matter bounce \cite{matterbounce} are all
investigated recently.

In this paper we will focus on the multiple fields DBI inflation
model. For single field DBI models, the effective four-dimensional
scalar field corresponds to the radial position of a brane in a
higher dimensional warped conical geometry with other angular degree
of freedom frozen for simplicity. However, if we consider the brane
can also move in the angular directions, more than one effective
scalars will turn on \cite{spininf}. Generally the Lagrangian of the
multifield DBI model can be expressed as $P(X^{IJ},\phi^K)$ with
$X^{IJ}=-\partial_{\mu}\phi^I\partial^{\mu}\phi^J/2$ , where
$I,J,K=1,2,\cdots,\mathcal{N}$ labels the $\mathcal{N}$ multiple
fields. And the trajectory of the background fields can be
decomposed into one adiabatic mode which is along the direction of
the trajectory and the other $(\mathcal{N}-1)$ entropic modes which
are orthogonal to the adiabatic direction. In contrast with the
single field model, in which the curvature perturbations on uniform
energy density hypersurfaces $\zeta$ are conserved after horizon
crossing, the curvature perturbations generally evolve in time even
on large scales in the multifield models. The reason of this effect
can be interpreted as due to the transfer between the adiabatic and
entropic modes \cite{transfer,RenauxPetel:2008gi,Ji:2009yw}, so if
there exists a large transfer from entropic modes to adiabatic mode,
then the final curvature perturbations are mostly of entropic
origin. Considering this effect, in this paper we calculate the
contributions from the interaction of four entropic perturbations
mediating one adiabatic perturbation to the trispectra in the
general multifield scenario.

This paper is organized as follows. In Sec.~II we firstly present
the background setup, solve the equation of motion for the
background fields, and secondly do the linear perturbations on the
background, derive the second order action, and calculate the
adiabatic and entropic power spectra in the massless limit. In first
subsection of the Sec. III we calculate the 4-point correlation
functions of the entropic perturbations $Q_s$ from the interaction
of four entropic modes mediating one adiabatic mode. In the second
subsection we further analyze the shape function of the trispectra
in two specific momenta configurations, one is the ``equilateral
configuration" another is the ``specialized planar configuration".
In Sec IV. we conclude our results.

\section{\label{section2}Review of generalized multifield model}
In this section, we firstly review the setup of the general
multifield model and the dynamics of the cosmological background,
then concentrate on the linear perturbation theory, including the
second order action and power spectra.
\subsection{Setup and Background}
In this subsection we briefly review the general multifield model
which was proposed by Langlois {\it et. al.} in~\cite{Langlois1},
where the Lagrangian is proposed to be of the form
$P(X^{IJ},\phi^K)$ with
$X^{IJ}\equiv-\frac{1}{2}\partial_{\mu}\phi^I\partial^{\mu}\phi^J$,
in particular, for the multifield DBI model the Lagrangian can be
reduced into \be{\label{lagrangian}}P(X^{IJ},
\phi^K)=-\frac{1}{f(\phi^I)}(\sqrt{\mathcal{D}}-1)-V(\phi^I)\;,~~~I=1,2,\cdots
\mathcal{N}\;,\ee where the determinant
$\mathcal{D}\equiv\det(\delta^{\mu}_{\nu}+fG_{IJ}\partial^{\mu}\phi^I\partial_{\nu}\phi^J)$,
$f(\phi^I)$ is the warping factor and $G_{IJ}$ is the field space
metric. As a specific example, such as the standard AdS throat, the
warp factor depends only on one of the fields and takes form of
$f(\phi_1)=\lambda/\phi^4_1$ (where $\lambda$ depends on the flux
numbers in specific string constructions).

Considering $\mathcal{N}$ scalar fields coupled with gravity
minimally, the action can be written as
\begin{eqnarray}
\label{DBI action}
S&=&\frac{1}{2}\int~d^4x~\sqrt{-g}\left[\mathcal{R}+2P(X^{IJ},\phi^K)\right]\;,
\end{eqnarray}
with $8\pi G=1$.

The energy-momentum tensor can be derived by varying
$P(X^{IJ},\phi^K)$ with respect to the metric $g_{\mu\nu}$
\be{\label{energy momentum tensor}}T^{\mu\nu}=Pg^{\mu\nu}+P_{\langle
IJ\rangle}\partial^{\mu}\phi^I\partial^{\nu}\phi^J\;,\ee where we
define \be{\label{PX}}P_{\langle
IJ\rangle}\equiv\frac{1}{2}\left(\frac{\partial P}{\partial
X^{IJ}}+\frac{\partial P}{\partial X^{JI}}\right)=P_{\langle
JI\rangle}\;.\ee

In the last part of this section, we will investigate the background
dynamics in the homogeneous and isotropic universe with the flat
Friedmann-Robertson-Walker metric
\be{\label{FRW}}ds^2=-dt^2+a^2(t)dx^idx^j\;,\ee where $a(t)$ is the
scale factor and $H=\dot a/a$ is the Hubble parameter. Thus, from
(\ref{energy momentum tensor}) we can see that the pressure is
simply $P$ and the energy density can be expressed as
\be{\label{energy dens}}\rho=2P_{\langle IJ\rangle}X^{IJ}-P\;,\ee
with $X^{IJ}=\frac{1}{2}\dot\phi^I\dot\phi^J$ and $\dot{}=d/dt$.

Following the assumption of metric (\ref{FRW}), the equation of
motion for scalar fields, the Friedmann equation and the continuity
equation will be reduced to
\begin{eqnarray}
\label{redu eom} 0&=&(P_{\langle IJ\rangle}+P_{\langle
IL\rangle,\langle
JK\rangle}\dot\phi^L\dot\phi^K)\ddot\phi^J+(3HP_{\langle
IJ\rangle}+P_{\langle
IJ\rangle,K}\dot\phi^K)\dot\phi^J-P_{,I}\;,\\
H^2&=&\frac{1}{3}(2P_{\langle IJ\rangle}X^{IJ}-P)\;,\\
\dot H&=&-X^{IJ}P_{\langle IJ\rangle}\;.
\end{eqnarray}

\subsection{Linear perturbation: Second order action and Power spectra}
In this subsection, we will briefly mention the derivation of the
second order action and derive the power spectra for the two fields
DBI model. Following the standard approach which was proposed in
\cite{maldacena}, one can get the rigorous second order action of
the general multifield inflation model
\begin{eqnarray}
\label{2nd order action1} S_{(2)}&=&\frac{1}{2}\int
dtd^3x~a^3\left[\left(P_{\langle IJ\rangle}+2P_{\langle
MJ\rangle,\langle IK\rangle}X^{MK}\right)\dot Q^I\dot Q^J-P_{\langle
IJ\rangle}h^{ij}\partial_iQ^I\partial_jQ^J\right.\nonumber\\
&&\left.-\mathcal{M}_{KL}Q^KQ^L+2\Omega_{KI}Q^K\dot Q^I\right]\;,
\end{eqnarray}
where the explicit form of mass matrix $\mathcal{M}_{KL}$ and mixing
matrix $\Omega_{KI}$ read
 \bea\label{mass matrix}
 \ma{M}_{KL}&=&-P_{,KL}+3X^{MN}P_{\la NK\ra}P_{\la ML\ra}+\f{1}{H}P_{\la
 NL\ra}\dot\phi^N\left[2P_{\la IJ\ra,K}X^{IJ}-P_{,K}\right]-\f{1}{H^2}X^{MN}\nonumber\\&&
 \times P_{\la NK\ra}P_{\la
  ML\ra}\left[X^{IJ}P_{\la IJ\ra}
 +2P_{\la IJ\ra,\la AB\ra}X^{IJ}X^{AB}\right]-\f{1}{a^3}\f{d}{dt}\l(\f{a^3}{H}
 P_{\la AK\ra}P_{\la LJ\ra}X^{AJ}\r)\;,\eea
 \be\label{mixing}
 \Omega_{KI}=\dot\phi^JP_{\la IJ\ra,K}-\f{2}{H}P_{\la LK\ra}P_{\la MJ\ra,\la
 NI\ra}X^{LN}X^{MJ}\;.\ee

In the derivation of (\ref{2nd order action1}), we used the ADM
metric \cite{ADM}
\be{\label{adm}}ds^2=-N^2dt^2+h_{ij}(dx^i+N^idt)(dx^j+N^jdt)\;,\ee
and chose the spatially flat gauge $h_{ij}=a^2\delta_{ij}$. In this
gauge, the spatial part of the metric remains unperturbed, thus the
physical degree of freedom are the perturbations of the multiple
scalar fields which are denoted as $Q^I$. Generally the matrix
(\ref{mass matrix}) is not diagonal, so $Q^I$ are not the canonical
quantities which can be promoted to operators through canonical
quantization procedure, so we must construct the canonical
quantities in the next step. Fortunately, we can achieve our aim
through decomposing $Q^I$ into one adiabatic mode which is along the
background trajectory in field space and $(\mathcal{N}-1)$ entropic
modes which are orthogonal to the trajectory, i.e. \be{\label{orth
decomp}}Q^I=Q^{\sigma}e^I_{\sigma}+Q^se^I_s\;,~~~s=2,\cdots,\mathcal{N}\;,\ee
where $e^I_{\sigma}$ is the unit base vector which is along the
trajectory, $e^I_s$ are the $(\mathcal{N}-1)$ unit base vectors
which are orthogonal to the trajectory, and they satisfy the
normalized and orthogonal relation
\be{\label{bases}}e^I_{\sigma}e_{\sigma I}=e^I_se_{sI}=1\;,
e^I_{\sigma}e_{sI}=0\;, e_{\sigma I}=G_{IJ}e^J_{\sigma}\;.\ee In
order to further simplify our calculation, we will assume the
straight line background trajectory and flat field space metric,
i.e., $\dot e^I_{\sigma}=\dot e^I_s=0$, and $G_{IJ}=\delta_{IJ}$.

Before doing the orthogonal decomposition, we need to stop here and
introduce the sound speed of perturbations in multifield DBI model.
As illustrated in \cite{Langlois1}, the remarkable result of the
action (\ref{lagrangian}) is that all perturbations propagate at the
same sound speed $c_s=\sqrt{1-fG_{IJ}\dot\phi^I\dot\phi^J}$.

After introducing the sound speed in multifield DBI model, we can
define a new field space metric as
\begin{eqnarray}
\label{diag metric} \tilde
G_{IJ}&=&\bot_{IJ}+\frac{1}{c^2_s}e_{\sigma I}e_{\sigma
J}\;,\\
\bot_{IJ}&=&G_{IJ}-e_{\sigma I}e_{\sigma J}\;,
\end{eqnarray}
where $\bot_{IJ}$ represents the projection to the entropy direction
and $e_{\sigma I}e_{\sigma J}/c^2_s$ represents the projection to
the adiabatic direction. Here we emphasize that the tilde metric is
diagonal, and using this metric we can reduce (\ref{2nd order
action1}) into\be{\label{2nd order action2}} S_{(2)}=\frac{1}{2}\int
dtd^3x~a^3\left[\frac{1}{c_s}\left(\tilde G_{IJ}\dot Q^I\dot Q^J
-c^2_s\frac{\tilde
G_{IJ}}{a^2}\partial_iQ^I\partial_iQ^J\right)-\tilde M_{IJ}Q^IQ^J
+2\frac{f_{,I}X}{c^3_s}\dot\phi_IQ^J\dot Q^I\right]\;.\ee

In order to gain some intuition, in what follows, we will restrict
ourselves to a two field model $(\phi_1,\phi_2)$, and it is
straightforward to generalize our analysis to any number of fields.
By virtue of (\ref{orth decomp}), $(\phi_1,\phi_2)$ are decomposed
into $(\sigma,s)$ with $\dot\sigma=\sqrt{2X}$ and $\dot s=0$. After
introducing three "slow variation parameters" as in standard slow
roll inflation
\begin{eqnarray}
\label{sr} \epsilon&=&-\frac{\dot H}{H^2}=\frac{X}{c_sH^2}\;,\\
\tilde\eta&=&\frac{\dot\epsilon}{\epsilon H}\;,\\
\tilde s&=&\frac{\dot c_s}{c_s H}\;,
\end{eqnarray}
one can reduce (\ref{2nd order action2}) further
\begin{eqnarray}
\label{2nd order action3} S_{(2)}&\simeq&\frac{1}{2}\int~d\eta
d^3x~\frac{1}{c_sH^2\eta^2}\left\{\frac{1}{c^2_s}\left[(Q'_{\sigma})^2-c^2_s(\partial_iQ^{\sigma})^2\right]
+\left[(Q'_s)^2-c^2_s(\partial_iQ^s)^2\right]\right\}\;,
\end{eqnarray}
where we change the cosmic time $t$ into comoving time $\eta=-1/aH$
($'=d/d\eta$) and drop the last two sub-leading terms in (\ref{2nd
order action2}) as in \cite{Langlois1}, because these two terms are
suppressed by the slow roll parameters which are defined in
(\ref{sr}). From (\ref{2nd order action3}), we can see that
$Q_{\sigma}$ and $Q_s$ are the canonical quantities, up to a
normalization factor, which should be quantized, and the propagation
speed of both the adiabatic mode and the entropy mode equal to
$c_s$.

Then we go to momentum space to do quantization, the Fourier mode of
$Q_{\sigma}$ and $Q_s$ can be quantized as
\begin{eqnarray}
\label{fourier} Q_{\sigma}(\eta,\bf k)&=&a_{\bf
k}u_k(\eta)+a^{\dagger}_{-\bf k}u^{\ast}_k(\eta)\;,\\
Q_s(\eta,\bf k)&=&b_{\bf k}v_k(\eta)+b^{\dagger}_{-\bf
k}v^{\ast}_k(\eta)\;,
\end{eqnarray}
where the creation and annihilation operators satisfy the standard
communication relation $\left[a({\bf k}),a^{\dagger}({\bf
k'})\right]=\left[b({\bf k}),b^{\dagger}({\bf
k'})\right]=(2\pi)^3\delta^3({\bf k-k'})$, and we choose the
Bunch-Davies vacuum
\begin{eqnarray}
\label{uv} u_k&=&\frac{H}{\sqrt{2k^3}}(1+ikc_s\eta)e^{-ikc_s\eta}\;,\\
v_k&=&\frac{H}{\sqrt{2k^3}c_s}(1+ikc_s\eta)e^{-ikc_s\eta}\;.
\end{eqnarray}

It is now straightforward to calculate the two point functions
\begin{eqnarray}
\label{two point} \left\langle Q_{\sigma}(\eta,{\bf
k}_1)Q_{\sigma}(\eta',{\bf k}_2)\right\rangle
&=&(2\pi)^3\delta^3({\bf k}_1+{\bf k}_2)F^>_{k_1}(\eta,\eta')\;,\\
\left\langle Q_{\sigma}(\eta',{\bf k}_2)Q_{\sigma}(\eta,{\bf
k}_1)\right\rangle
&=&(2\pi)^3\delta^3({\bf k}_1+{\bf k}_2)F^<_{k_1}(\eta,\eta')\;,\\
\left\langle Q_{s}(\eta,{\bf k}_1)Q_{s}(\eta',{\bf
k}_2)\right\rangle
&=&(2\pi)^3\delta^3({\bf k}_1+{\bf k}_2)G^>_{k_1}(\eta,\eta')\;,\\
\left\langle Q_{s}(\eta',{\bf k}_2)Q_{s}(\eta,{\bf
k}_1)\right\rangle &=&(2\pi)^3\delta^3({\bf k}_1+{\bf
k}_2)G^<_{k_1}(\eta,\eta')\;,
\end{eqnarray}
where we set $\eta>\eta'$, and the Wightman functions for adiabatic
and entropy modes read
\begin{eqnarray}
\label{green}
F^>_k(\eta,\eta')&=&u_k(\eta)u^{\ast}_k(\eta')\;,~~F^<_k(\eta,\eta')=u^{\ast}_k(\eta)u_k(\eta')\;,\\
G^>_k(\eta,\eta')&=&v_k(\eta)v^{\ast}_k(\eta')\;,~~G^<_k(\eta,\eta')=v^{\ast}_k(\eta)v_k(\eta')\;.
\end{eqnarray}
Then the power spectra of $Q_{\sigma}$ and $Q_{s}$ are
\begin{eqnarray}
\label{power sigma s}
P^{\sigma}_k&=&|Q_{\sigma^{\ast}}|^2=\frac{H_{\ast}^2}{2k^3}\;,\\
P^{s}_k&=&|Q_{s^{\ast}}|^2=\frac{H_{\ast}^2}{2k^3c^2_s}\;,
\end{eqnarray}
where the subscript $\ast$ indicates that the corresponding quantity
is evaluated at the sound horizon crossing $kc_s=aH$.

In single field model, the curvature perturbation $\ma{R}$
($\ma{R}=-\zeta$) remains constant in the large scale limit due to
the local energy conservation. However, in multifield model, the
adiabatic perturbations can not produce the entropic perturbations,
but the entropic perturbations can source the curvature
perturbations on the large scales. Generally, the time dependence of
the adiabatic and entropic perturbations on the superhorizon scales
are always described by
 \be\label{superh1}\dot\ma{R}=\alpha H
 \ma{S}\;,\qquad\dot\ma{S}=\beta H\ma{S}\;,\ee
 with
 \bea\label{adi entr}
 \ma{R}\equiv\f{H}{\dot\s}Q_{\s}\;,\qquad \ma{S}\equiv c_s\f{H}{\dot{\s}}Q_s\;,\eea
and $\a$, $\b$ are time dependent dimensionless functions. After
performing the time integration for (\ref{superh1}), one can get the
general form of the transfer matrix which relate the curvature and
entropic perturbations generated when a given mode crosses the sound
horizon at time $t_{\ast}$ to those at some later time $t$ (on
superhorizon scales)\cite{Langlois1,transfer,2fieldsob}
  \eq{
        \lrp{\begin{array}{c}
               \mathcal{R} \\
               \mathcal{S}
             \end{array}
         } = \lrp{ \begin{array}{cc}
                     1 & T_{\mathcal{R}\mathcal{S}} \\
                     0 & T_{\mathcal{S}\mathcal{S}}
                   \end{array}
          } \lrp{\begin{array}{c}
               \mathcal{R} \\
               \mathcal{S}
             \end{array}
         }_{\ast} \,,\label{transf}
    }
where
 \eq{
        T_{\mathcal{S}\mathcal{S}}(t,t_{\ast}) = exp\left\{{\int_{t_{\ast}}^t dt'\, \beta(t')H(t')
        }\right\}\,,\qquad T_{\mathcal{R}\mathcal{S}}(t,t_{\ast}) = \int_{t_{\ast}}^t
        dt'\, \alpha(t') T_{\mathcal{S}\mathcal{S}}(t',t_{\ast})
        H(t') \,,
    }

Substitute (\ref{adi entr}) into (\ref{transf}), one can obtain the
relationship between the $\zeta(t)$ and $Q_{i}(t_{\ast})$ with
$i=\s, s$
\be{\label{zeta}}\zeta(t)=-\mathcal{A}_{\sigma}Q_{\sigma}(t_{\ast})-\mathcal{A}_sQ_{s}(t_{\ast})\;,\ee
where
 \be{\label{Asigma
As}}\mathcal{A}_{\sigma}=\left(\frac{H}{\dot\sigma}\right)_{\ast},\qquad
\mathcal{A}_{s}=T_{RS}(t,t_{\ast})\left(\frac{c_sH}{\dot\sigma}\right)_{\ast}\;,\ee
and $T_{RS}(t,t_{\ast})$ is the transfer coefficient which reflects
the transfer between the adiabatic and entropic modes.

If $T_{RS}(t,t_{\ast})\gg1$,
$\zeta(t)\simeq-\mathcal{A}_s(t,t_{\ast})Q_s(t_{\ast})$, i.e., the
curvature perturbations on superhorizon scales are mainly
transferred from the entropic perturbations. Thus the late time
power spectrum of $\zeta(t)$ becomes
\be{\label{power}}P^{\zeta}_k(t)\simeq
T_{RS}^2(t,t_{\ast})\frac{c_s}{2\epsilon}|Q_{s^{\ast}}|^2=T_{RS}^2(t,t_{\ast})\frac{H^2_{\ast}}{4\epsilon
c_sk^3}\;.\ee

\section{\label{section3}Trispectra from ``scalar-exchanging" interaction}
In this section, we will derive the general form of the 4-point
correlation functions and plot the shape function (momentum
dependence) for two special momenta configurations.
\subsection{\label{3.1}Four point correlation functions}
In the first subsection, we will calculate the 4-point correlation
functions for the entropic perturbations $Q_s$ through the
``scalar-exchanging" interaction. For this purpose, we firstly
listed the third order action which has be derived in
\cite{Langlois1}
\begin{eqnarray}
\label{3rd order action} S^{{\rm main}}_{(3)}&=&\int~d\eta
d^3x~\left\{\frac{1}{2H^2\eta^2c^5_s\sigma'}\left[(Q'_{\sigma})^3-c^2_sQ'_{\sigma}(\nabla
Q_{\sigma})^2\right]\right.\nonumber\\
&&\left.+\frac{1}{2H^2\eta^2c^3_s\sigma'}\left[
Q'_{\sigma}(Q'_s)^2+c^2_sQ'_{\sigma}(\nabla Q_s)^2-2c^2_sQ'_s\nabla
Q_s\nabla Q_{\sigma}\right]\right\}\;.
\end{eqnarray}

As proved in \cite{dbi tri}, up to the third order, the Hamiltonian
equals to the opposite Lagrangian $H^I_{(3)}=-L^I_{(3)}$, where the
supper script ``$I$" denotes for the interaction picture
\begin{eqnarray}
\label{HI}
H^I_1(\eta)&=&\frac{-1}{2H^2\eta^2c^5_s\sigma'}\left[\prod^3_{i=1}\int\frac{d^3k_i}{(2\pi)^3}\right]
(2\pi)^3\delta^3({\bf k}_{123})Q'_{\sigma}(\eta,{\bf
k}_1)Q'_{\sigma}(\eta,{\bf k}_2)Q'_{\sigma}(\eta,{\bf k}_3)\;,\\
H^I_2(\eta)&=&\frac{-1}{2H^2\eta^2c^3_s\sigma'}\left[\prod^3_{i=1}\int\frac{d^3k_i}{(2\pi)^3}\right]
(2\pi)^3\delta^3({\bf k}_{123})({\bf k}_2\cdot{\bf
k}_3)Q'_{\sigma}(\eta,{\bf k}_1)Q_{\sigma}(\eta,{\bf
k}_2)Q_{\sigma}(\eta,{\bf k}_3)\;,\\
H^I_3(\eta)&=&\frac{-1}{2H^2\eta^2c^3_s\sigma'}\left[\prod^3_{i=1}\int\frac{d^3k_i}{(2\pi)^3}\right]
(2\pi)^3\delta^3({\bf k}_{123})Q'_{\sigma}(\eta,{\bf
k}_1)Q'_{s}(\eta,{\bf k}_2)Q'_{s}(\eta,{\bf k}_3)\;,\\
H^I_4(\eta)&=&\frac{1}{2H^2\eta^2c_s\sigma'}\left[\prod^3_{i=1}\int\frac{d^3k_i}{(2\pi)^3}\right]
(2\pi)^3\delta^3({\bf k}_{123})({\bf k}_2\cdot{\bf
k}_3)Q'_{\sigma}(\eta,{\bf k}_1)Q_{s}(\eta,{\bf
k}_2)Q_{s}(\eta,{\bf k}_3)\;,\\
H^I_5(\eta)&=&\frac{-1}{~~H^2\eta^2c_s\sigma'}\left[\prod^3_{i=1}\int\frac{d^3k_i}{(2\pi)^3}\right]
(2\pi)^3\delta^3({\bf k}_{123})({\bf k}_2\cdot{\bf
k}_3)Q'_{s}(\eta,{\bf k}_1)Q_{s}(\eta,{\bf k}_2)Q_{\sigma}(\eta,{\bf
k}_3)\;.
\end{eqnarray}

\begin{figure}[h]
    \centering
    \begin{minipage}{0.8\textwidth}
    \centering
        \begin{minipage}{0.3\textwidth}
            \centering
            \includegraphics[width=3.5cm]{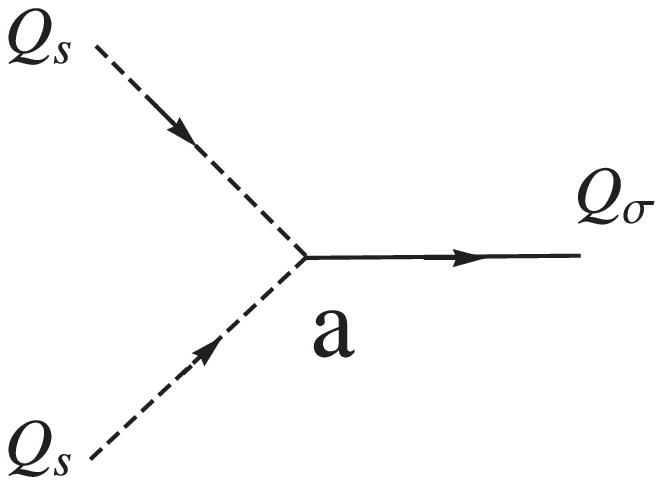}
        \end{minipage}
        \begin{minipage}{0.3\textwidth}
            \centering
            \includegraphics[width=3.5cm]{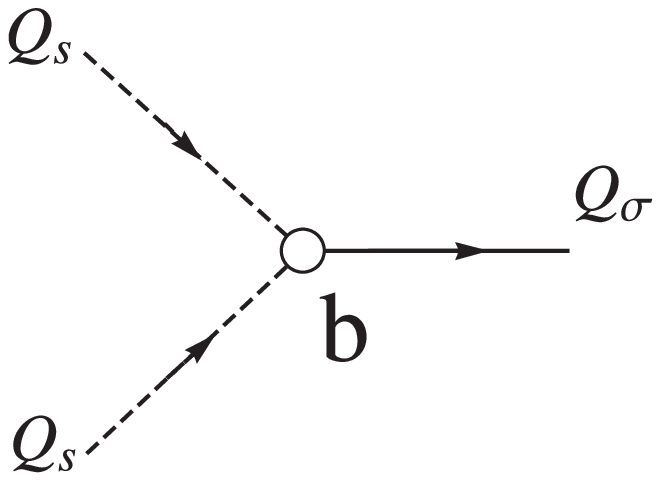}
        \end{minipage}
        \begin{minipage}{0.3\textwidth}
            \centering
            \includegraphics[width=3.5cm]{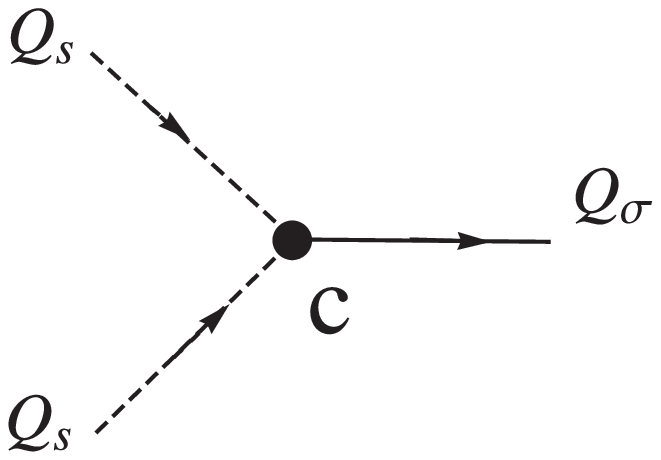}
        \end{minipage}
    \end{minipage}
    \caption{Diagrammatic representation of the 3-point
     vertices: dashed line denotes for the entropic mode $Q_s$,solid
             line for the adiabatic mode $Q_{\sigma}$, and vertex $a$ presents for the
             interaction $H^I_3$, vertex $b$ for $H^I_4$, vertex $c$
             for $H^I_5$.}
    \label{ver}
    \end{figure}

In what follows, we will still concentrate on the $T_{RS}\gg1$ case.
For this one, the main contributions to trispectra comes from the
vertices with two external entropic legs $H^I_3$, $H^I_4$, and
$H^I_5$ (see Fig. \ref{ver}), because on the large scales the
entropic perturbations source the curvature perturbations, i.e., we
can approximately take $H^I\simeq H^I_3+H^I_4+H^I_5$.

\begin{figure}[h]
\centering
\includegraphics[width=5cm]{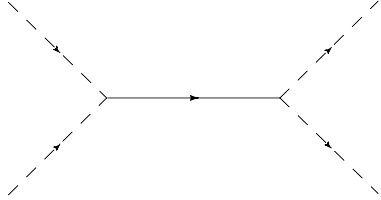}
\caption{\label{33}This figure illustrates the interaction between
four entropic modes through mediating one adiabatic mode.}
\end{figure}

Now, we are ready for calculating the 4-point correlation functions
of the entropic perturbations $Q_s$. The ``scalar-exchanging"
interaction can be illustrated diagrammatically in Fig. \ref{33}, in
which we merely plot one of the nine similar diagrams. Since there
exist three different 3-point vertices, as shown in Fig. \ref{ver},
the 4-point correlation functions of the entropic modes can be
expressed as{\footnote{The in-in formalism \cite{in-in} which is
often used in the literature in terms of a commutator form, is
equivalent to the form (\ref{4pt3terms}) presented here.}}

 \be\label{Qtri}
 \langle Q_s^4(\y_{\ast}) \rangle=\sum_{i,j=3}^5\langle Q_s^4(\y_{\ast})
 \rangle_{ij}\;,\ee
 with
 \bea
 \label{4pt3terms}
 \langle Q_s^4(\y_{\ast}) \rangle_{ij} &=& \langle 0| \left[{\bar T}
e^{i\int_{\y_0}^{\y_{\ast}} d\y'
  H_I(\y')} \right]
  Q_s(\bp_1,\y_{\ast}) Q_s(\bp_2,\y_{\ast}) Q_s(\bp_3,\y_{\ast})
  Q_s(\bp_4,\y_{\ast})
  \left[ T e^{-i\int_{\y_0}^{\y_{\ast}} d\y' H_I(\y')} \right] |0
  \rangle\nonumber
\\
&\simeq& \int_{\y_0}^{\y_{\ast}} d\y' \int_{\y_0}^{\y_{\ast}} d\y''
~\langle 0| ~H_i(\y') ~Q_s^4(\y_{\ast}) ~H_j(\y'') ~|0 \rangle
\nonumber \\
&-& \int_{\y_0}^{\y_{\ast}} d\y' \int_{\y_0}^{\y'} d\y'' ~\langle 0|
~H_j(\y'') ~H_i(\y') ~Q_s^4(\y_{\ast}) ~|0\rangle
\nonumber \\
&-& \int_{\y_0}^{\y_{\ast}} d\y' \int_{\y_0}^{\y'} d\y'' ~\langle 0|
~Q_s^4(\y_{\ast}) ~H_i(\y') ~H_j(\y'') ~|0\rangle ~. \eea

After some straightforward but lengthy calculations we can obtain
the analytic expressions of the 4-point correlation functions of the
entropic modes
 \bea
 \la
 Q_s^4(\y_{\ast})\ra_{33}&=&\f{(2\pi)^3\d^3(\sum_{i=1}^4 \bp_i)}{\prod_{i=1}^4
 p_i^3}\f{H^6}{2^6\e c_s^9}\nonumber\\
 &&\times\l\{p_{12}\l(\prod_{i=1}^4p_i^2\r)\l[\f{2(10q_1^2+5q_1q_2+q_2^2)}{q_1^3K^5}
 +\f{1}{q_1^3q_3^3}\r]+23~{\rm perm.}\r\}\;,
 \eea
 \bea
 \la
 Q_s^4(\y_{\ast})\ra_{44}&=&\f{(2\pi)^3\d^3(\sum_{i=1}^4 \bp_i)}{\prod_{i=1}^4
 p_i^3}\f{H^6}{2^8\e c_s^9}p_{12}(\bp_1\cdot\bp_2)(\bp_3\cdot\bp_4)
 \times\l\{F(p_1,p_2,q_3)F(p_3,p_4,q_1)\r.\nonumber\\
 &&\l.+2G(p_1,p_2,p_4,p_3,q_1)\r\}+{\rm 23~perm.}\;,
 \eea
 \bea
 \la Q_s^4(\y_{\ast})\ra_{55}&=&\f{(2\pi)^3\d^3(\sum_{i=1}^4
 \bp_i)}{\prod_{i=1}^4 p_i^3}\f{H^6}{2^6\e
 c_s^9}\f{-p_1^2p_3^2(\bp_2\cdot\bp_{12})(\bp_4\cdot\bp_{12})}{p_{12}^3}
 \times\l\{F(p_{12},p_4,q_1)F(p_{12},p_2,q_3)\r.\nonumber\\
 &&\l.+2G(-p_{12},p_2,p_4,p_{12},q_1)\r\}
 +{\rm 23~perm.}\;,
 \eea
 \bea
 \la Q_s^4(\y_{\ast})\ra_{34}&=&\f{(2\pi)^3\d^3(\sum_{i=1}^4
 \bp_i)}{\prod_{i=1}^4 p_i^3}\f{H^6}{2^7\e
 c_s^9}p_{12}p_1^2p_2^2(\bp_3\cdot\bp_4)\l\{\f{F(p_3,p_4,q_1)}{q_3^3}
 +E(p_3,p_4,q_1)\r\}\nonumber\\
 &&+{\rm 23~perm.}\;,
 \eea
 \bea
 \la Q_s^4(\y_{\ast})\ra_{43}&=&\f{(2\pi)^3\d^3(\sum_{i=1}^4
 \bp_i)}{\prod_{i=1}^4 p_i^3}\f{H^6}{2^7\e
 c_s^9}p_{12}p_3^2p_4^2(\bp_1\cdot\bp_2)\l\{\f{F(p_1,p_2,q_3)}{q_1^3}
 +H(p_1,p_2,q_1)\r\}\nonumber\\
 &&+{\rm 23~perm.}\;,
 \eea
 \bea
 \la Q_s^4(\y_{\ast})\ra_{35}&=&\f{(2\pi)^3\d^3(\sum_{i=1}^4
 \bp_i)}{\prod_{i=1}^4 p_i^3}\f{H^6}{2^6\e
 c_s^9}\f{-p_1^2p_2^2p_3^2(\bp_4\cdot\bp_{12})}{p_{12}}
 \l\{\f{F(p_{12},p_4,q_1)}{q_3^3}+E(p_{12},p_4,q_1)\r\}\nonumber\\
 &&+{\rm 23~perm.}\;,
 \eea
 \bea
 \la Q_s^4(\y_{\ast})\ra_{53}&=&\f{(2\pi)^3\d^3(\sum_{i=1}^4
 \bp_i)}{\prod_{i=1}^4 p_i^3}\f{H^6}{2^6\e
 c_s^9}\f{p_3^2p_4^2p_1^2(\bp_2\cdot\bp_{12})}{p_{12}}\l\{\f{F(p_{12},p_2,q_3)}{q_1^3}
 +H(-p_{12},p_2,q_1)\r\}\nonumber\\
 &&+{\rm
 23~perm.}\;,
 \eea
 \bea
 \la Q_s^4(\y_{\ast})\ra_{45}&=&\f{(2\pi)^3\d^3(\sum_{i=1}^4
 \bp_i)}{\prod_{i=1}^4 p_i^3}\f{H^6}{2^7\e
 c_s^9}\f{-p_3^2(\bp_1\cdot\bp_2)(\bp_4\cdot\bp_{12})}{p_{12}}
 \times\l\{ F(p_1,p_2,q_3)F(p_{12},p_4,q_1)\r.\nonumber\\
 &&\l.+2G(p_1,p_2,p_4,p_{12},q_1)\r\}
 +{\rm 23~perm.}\;,
 \eea
 \bea
 \la Q_s^4(\y_{\ast})\ra_{54}&=&\f{(2\pi)^3\d^3(\sum_{i=1}^4
 \bp_i)}{\prod_{i=1}^4 p_i^3}\f{H^6}{2^7\e
 c_s^9}\f{p_1^2(\bp_3\cdot\bp_4)(\bp_2\cdot\bp_{12})}{p_{12}}
 \times\l\{F(p_3,p_4,q_1)F(p_{12},p_2,q_3)\r.\nonumber\\
 &&\l.+2G(p_2,-p_{12},p_4,p_3,q_1)\r\}+{\rm
 23~perm.}\;,
 \eea
 with
 \be\label{q} q_1\equiv p_3+p_4+p_{12}\;,\qquad q_2\equiv p_1+p_2-p_{12}\;,\qquad q_3\equiv
 p_1+p_2+p_{12}\;,\qquad K\equiv \sum_{i=1}^4p_i\;,\ee
 \be
 \label{F}
 F(p_i,p_j,q_k)\equiv\f{2p_ip_j+(p_i+p_j)q_k+q_k^2}{q_k^3}\;,
 \ee
 \bea
 E(p_i,p_j,q_k)&\equiv&\f{4p_ip_j}{K^3q_k^3}
 +\f{2(p_i+p_j)}{K^3q_k^2}+\f{12p_ip_j}{K^4q_k^2}+\f{2}{K^3q_k}+\f{6(p_i+p_j)}{K^4q_k}
 +\f{24p_ip_j}{K^5q_k}\;,
 \eea
 \bea
 H(p_i,p_j,q_k)&\equiv&\f{4p_ip_j}{K^3q_k^3}
 +\f{2}{Kq_k^3}+\f{2}{K^2q_k^2}+\f{2}{K^3q_k}+\f{2(p_i+p_j)}{K^2q_k^3}+\f{4(p_i+p_j)}{K^3q_k^2}
 +\f{12p_ip_j}{K^4q_k^2}\nonumber\\
 &&+\f{6(p_i+p_j)}{K^4q_k}+\f{24p_ip_j}{K^5q_k}\;,
 \eea
 \bea
 G(p_i,p_j,p_l,p_m,q_n)&\equiv&\nonumber\\
 &&\f{2p_{m}p_l}{Kq_n^3}
 +\f{2p_{m}p_l(p_i+p_j)}{K^2q_n^3}+\f{4p_ip_jp_lp_{m}}{K^3q_n^3}+\f{p_{m}+p_l}{Kq_n^2}
 +\f{(p_i+p_j)(p_l+p_{m})+2p_lp_{m}}{K^2q_n^2}\nonumber\\
 &&+\f{2p_ip_j(p_{m}+p_l)+4p_{m}p_l(p_i+p_j)}{K^3q_n^2}+\f{12p_ip_jp_lp_{m}}{K^4q_n^2}
 +\f{1}{Kq_n}+\f{p_i+p_j+p_l+p_{m}}{K^2q_n}\nonumber\\
 &&+\f{2p_lp_{m}+2(p_i+p_j)(p_l+p_{m})+2p_ip_j}{K^3q_n}+\f{6\l[p_ip_j(p_l+p_{m})+p_lp_{m}(p_i+p_j)\r]}{K^4q_n}\nonumber\\
 &&+\f{24p_ip_jp_lp_{m}}{K^5q_n}\;,
 \eea
where $``{\rm 23~perm.}"$ denotes for the $23$ permutations between
the momenta $\bp_1,\bp_2,\bp_3$ and $\bp_4$.

Finally, we can get the general form of trispectra for the curvature
perturbations $\z$, by virtue of (\ref{zeta}) and (\ref{Asigma As})
 \bea
 \label{trispectra}
 \la\zeta^4(\y,\bp_1,\bp_2,\bp_3,\bp_4)\ra&\simeq&\mathcal{A}^4_s(\y,\y_{\ast})_s\la
 Q_s^4(\y_{\ast})\ra
 =T^4_{RS}(\y,\y_{\ast})\left(\frac{c_sH}{\dot\sigma}\right)^4_{\ast}\la
 Q_s^4(\y_{\ast})\ra\nonumber\\
 &\simeq&\f{T_{RS}^4(\y,\y_{\ast})c_s^2}{4\e^2}\la
 Q_s^4(\y_{\ast})\ra\nonumber\\
 &=&\f{(2\pi)^3H^6T_{RS}^4(\y,\y_{\ast})\d^3(\sum_{i=1}^4
 \bp_i)}{2^8\e^3c_s^7}\ma{A}(\bp_1,\bp_2,\bp_3,\bp_4)\;,
 \eea
 where in the second line we used the ``slow variation parameters" defined in (\ref{sr}), and the
 function $\ma{A}(\bp_1,\bp_2,\bp_3,\bp_4)$ defined in the third line is called shape
 function, which we will analyze numerically in the next subsection.

\subsection{\label{3.2}Shapes of the trispectra}
As we have obtained the general form of the trispectra, in this
subsection, we will turn to plot the shape diagrams for the
equilateral configuration with ($p_1=p_2=p_3=p_4$) and the
``specialized planar" configuration with ($p_3=p_4=p_{12}$).

Before the discussion of the shape functions, we note that the
number of the independent arguments for the trispectra are six. In
this paper, we choose six independent momenta $p_1, p_2, p_3, p_4,
p_{12}, p_{14}$, and one can also choose four independent momenta
and two angles. In order for these momenta to form a tetrahedron
(see Fig. \ref{tet}), two conditions must be satisfied
\cite{Chen:2009bc}:

\begin{figure}
    \centering
    \begin{minipage}{0.9\textwidth}
    \centering
        \begin{minipage}{0.45\textwidth}
        \centering
            \begin{minipage}{0.9\textwidth}
            \centering
            \includegraphics[width=5cm]{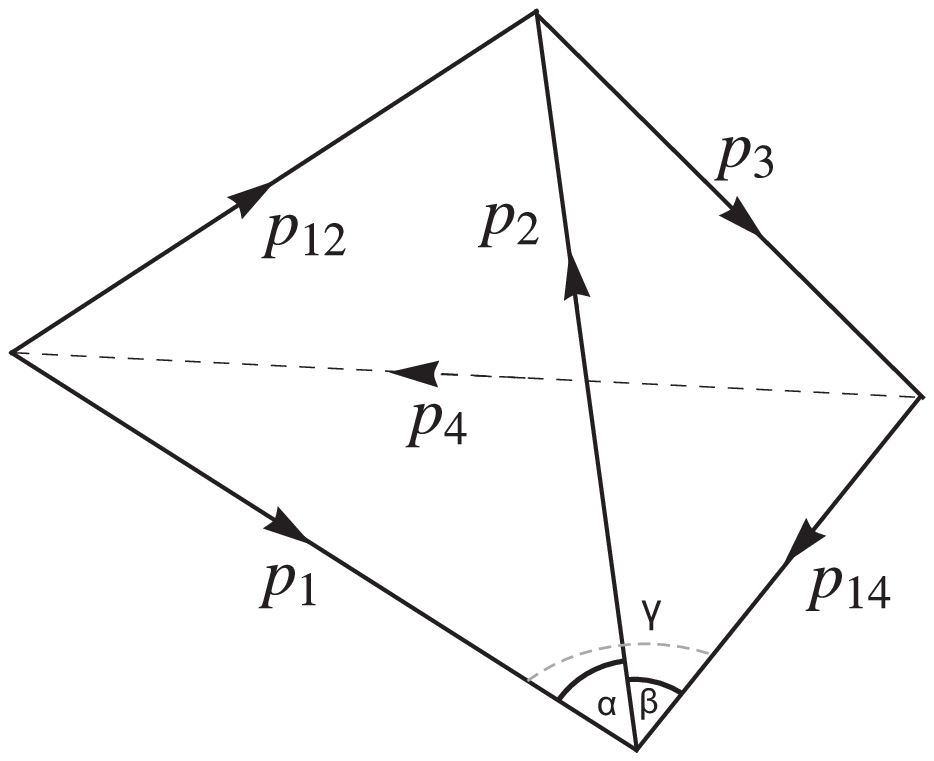}
            \caption{Tetrahedron configuration: momenta $(\bp_1,\bp_2,\bp_3,\bp_4)$ do not lie in
            the same plane, six momenta $(p_1,p_2,p_3,p_4,p_{12},p_{14})$ form a tetrahedron.}
            \label{tet}
            \end{minipage}
        \end{minipage}
    \begin{minipage}{0.45\textwidth}
    \centering
        \begin{minipage}{0.9\textwidth}
        \centering
        \includegraphics[width=5.2cm]{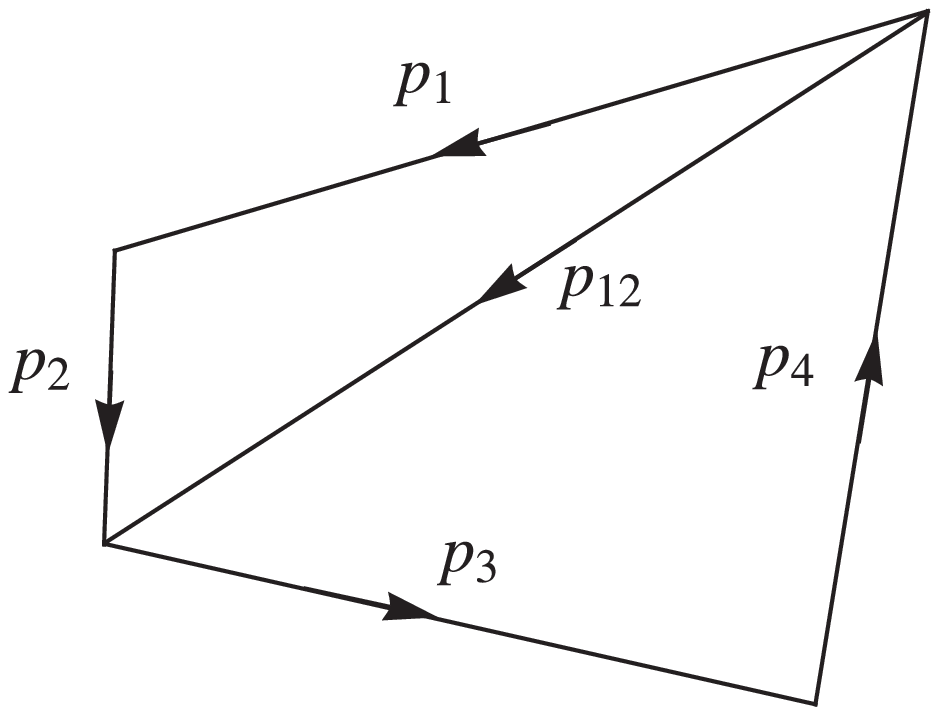}
         \caption{Planar configuration: momenta $(\bp_1,\bp_2,\bp_3,\bp_4)$ lie in the same plane,
         and form a planar quadrangle.}
            \label{qua}
            \end{minipage}
        \end{minipage}
         \end{minipage}
         \end{figure}

First, we define three angles at one vertex (see Fig. \ref{tet})
 \bea
 \label{alpha}\cos(\a)&=&\f{p_1^2+p_2^2-p_{12}^2}{2p_1p_2}\;,\\
 \label{beta}\cos(\b)&=&\f{p_2^2+p_{14}^2-p_3^3}{2p_2p_{14}}\;,\\
 \label{gamma}\cos(\gamma)&=&\f{p_1^2+p_{14}^2-p_4^2}{2p_1p_{14}}\;,
 \eea
where these three angles should satisfy
 $\cos(\a-\b)\geq\cos(\gamma)\geq\cos(\a+\b)$. This inequality is
 equivalent to
 \be
 \label{vol cond}
 1-\cos^2(\a)-\cos^2(\b)-\cos^2(\gamma)+2\cos(\a)\cos(\b)\cos(\gamma)\geq0\;,
 \ee
 where we can take the equal sign when the tetrahedron reduces to a
 planar quadrangle.

 Secondly, the six momenta should also satisfy all the triangle
 inequalities
 \bea
 \label{triangle ineq}
  p_1+p_4>p_{14}~,\quad   p_1+p_2>p_{12}~,\quad
  p_2+p_3>p_{14}~,\nonumber\\
  p_1+p_{14}>p_4~,\quad   p_1+p_{12}>p_2~,\quad
  p_2+p_{14}>p_3~,\nonumber\\
  p_4+p_{14}>p_1~,\quad   p_2+p_{12}>p_1~,\quad
  p_3+p_{14}>p_2~,
 \eea
 and the last triangle inequality involving ($p_3,p_4,p_{12}$) is
 always satisfied given (\ref{vol cond}) and (\ref{triangle ineq}).

\begin{figure}
    \centering
    \begin{minipage}{0.9\textwidth}
    \centering
        \begin{minipage}{0.45\textwidth}
        \centering
            \begin{minipage}{0.9\textwidth}
            \centering
            \includegraphics[width=5cm]{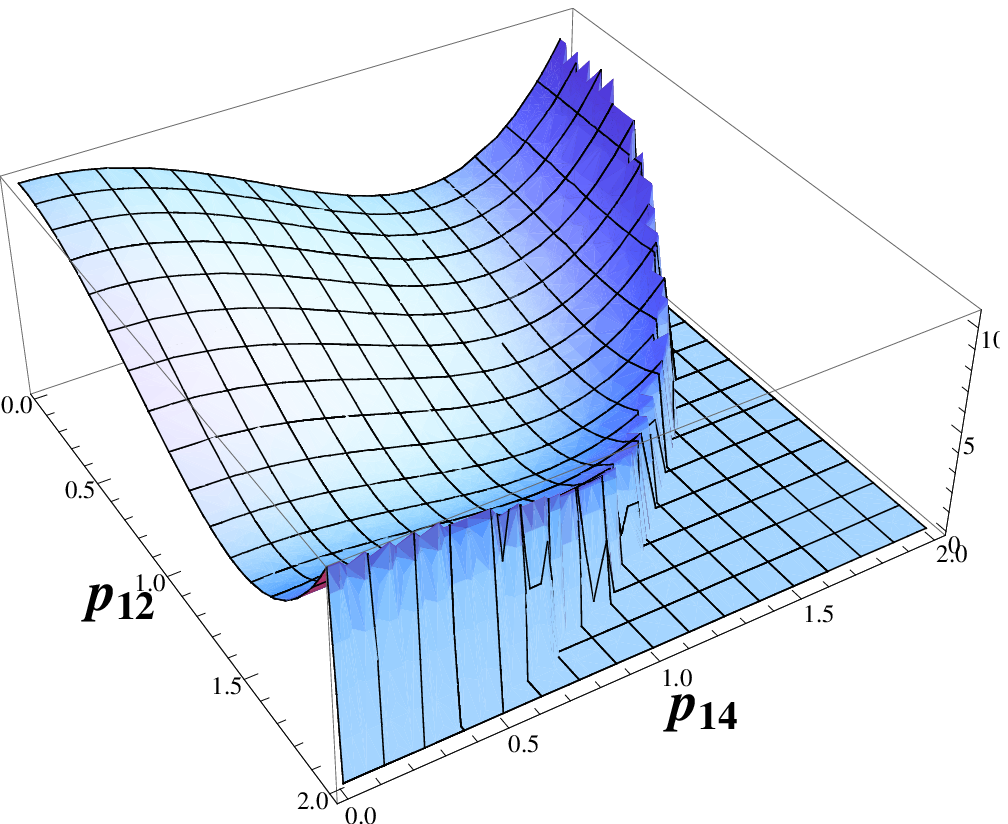}
            \caption{Shape of the equilateral configuration $\ma{A}(p_{12},p_{14})$:
            in this configuration we set ($p_1=p_2=p_3=p_4=1$).}
            \label{eq}
            \end{minipage}
        \end{minipage}
    \begin{minipage}{0.45\textwidth}
    \centering
        \begin{minipage}{0.9\textwidth}
        \centering
        \includegraphics[width=5cm]{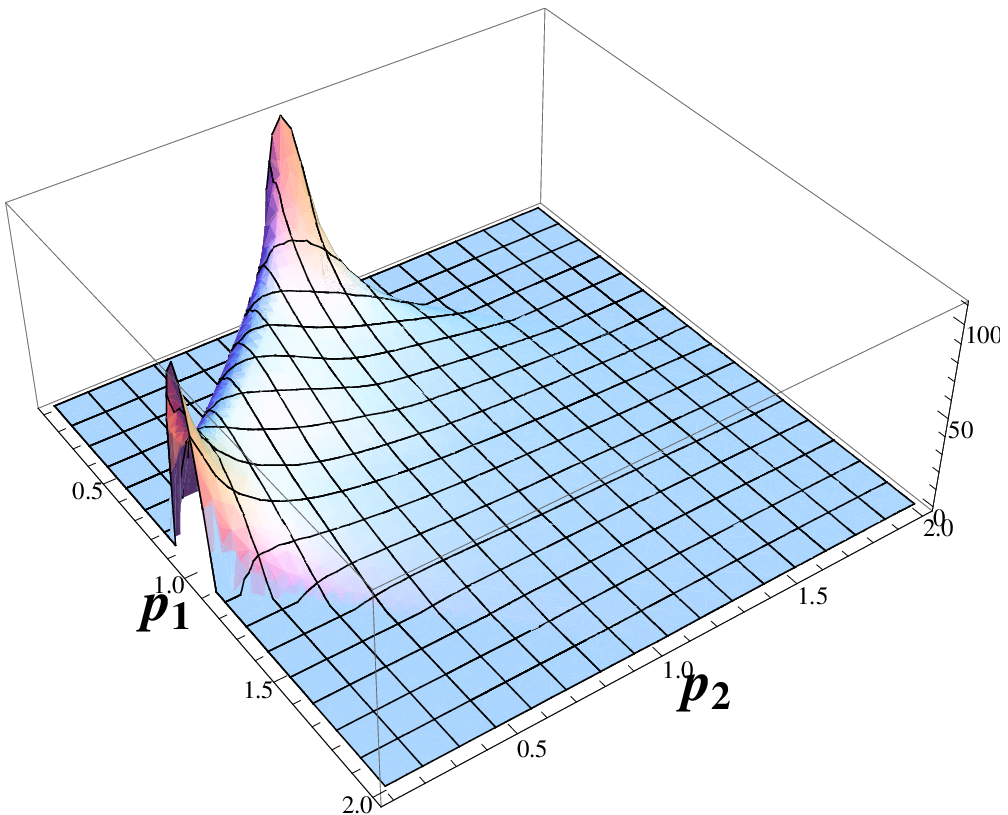}
         \caption{Shape of the ``specialized planar" configuration $\ma{A}(p_{1},p_{2})$:
         in this configuration we set ($p_3=p_4=p_{12}=1$).}
            \label{pl}
            \end{minipage}
        \end{minipage}
         \end{minipage}
         \end{figure}

 After discussing the two conditions which should be satisfied, now
we will plot the shape function
$\ma{A}(p_1,p_2,p_3,p_4,p_{12},p_{14})$ which is defined in
(\ref{trispectra}). The first momenta configuration which we are
interested in is called the ``equilateral configuration"
($p_1=p_2=p_3=p_4=1$). In this configuration, we plot the shape
function $\ma{A}(p_{12},p_{14})$ versus $p_{12}$ and $p_{14}$ (see
Fig. \ref{eq}). And one can easily see from Fig. \ref{eq}, that the
amplitude of the shape function blows up on the circle with radius
$2p_1$, which represents for the limit with $\bp_1\cdot\bp_{13}=0$
($p_{13}=0$).

In the second case, we consider a specialized planar momenta
configuration, i.e., the quadrangle with $p_3=p_4=p_{12}=1$ (see
Fig. \ref{qua}). As we have said, in the planar limit, (\ref{vol
cond}) takes the equal sign, so we can obtain $p_{14}$ by solving
(\ref{vol cond})
 \be
 \label{p14}
   p_{14}= \frac{\sqrt{p_1^2 \left(-p_{4}^2+p_3^2+p_{12}^2\right)\pm p_{s1}^2 p_{s2}^2+p_{12}^2 p_{2}^2+p_{12}^2 p_{4}^2+p_{2}^2
   p_{4}^2-p_{2}^2 p_3^2-p_{12}^4+p_3^2 p_{12}^2}}{\sqrt{2} p_{12}}~,
 \ee
where $p_{s1}$ and $p_{s2}$ are defined as
 \bea &  p_{s1}^2\equiv
 2\sqrt{(p_1 p_{12}+{\bf p}_1 \cdot {\bf p}_{12})(p_1 p_{12}-{\bf p}_1 \cdot
 {\bf
  p}_{12})}~,\nonumber\\ &
  p_{s2}^2\equiv 2\sqrt{(p_3 p_{12}+{\bf p}_3 \cdot {\bf p}_{12})(p_3
  p_{12}-{\bf p}_3 \cdot {\bf
  p}_{12})}~.
 \eea

\begin{figure}[h]
\centering
\includegraphics[width=5.5cm]{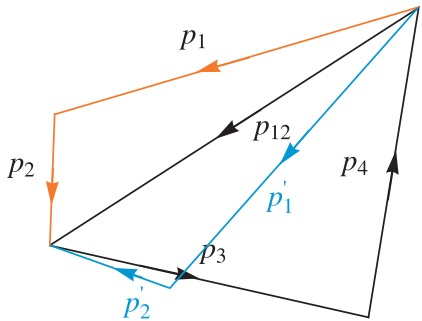}
\caption{\label{dual}This figure illustrates the two dual
quadrangles with the same absolute value of momenta
($p_1,p_2,p_3,p_4,p_{12}$): the orange one corresponds to the $+$
solution and the blue one to the $-$ solution.}
\end{figure}

As pointed out in \cite{Chen:2009bc}, the $-$ solution (blue) and
$+$ solution (orange) in fact dual with each other (see Fig.
\ref{dual}), so we can choose arbitrary one to discuss, in the
following we take the the $+$ solution.

After setting $p_3=p_4=p_{12}=1$ and solving $p_{14}$, the two
independent arguments of the shape function are $p_1$ and $p_2$, so,
in this ``specialized planar configuration", we plot the shape
function $\ma{A}(p_1,p_2)$ versus $p_1$ and $p_2$ (see Fig.
\ref{pl}). From Fig. \ref{pl}, one can see that the shape function
was highly-peaked at the ``squeezed limit" ($p_1, p_2\rightarrow0$).

\section{Conclusion}
In this paper, we investigated the primordial trispectra produced by
the ``scalar-exchanging" interaction of the general multifield DBI
inflationary model. In section \ref{section2}, we studied the power
spectra of the entropic modes and adiabatic modes without
considering the mass term and the mixing term $\la
Q_sQ_{\sigma}\ra$. In contrast with the single field model, the
curvature perturbations generally evolve in time on the large scales
in the multifield scenario, because the entropic modes can source
the curvature perturbations on the super horizon scales, which can
be described by the transfer coefficient $T_{RS}$. So, if the
transfer process is strong ($T_{RS}\gg1$), the late time curvature
perturbations will be mostly of the entropic origin.

Given that reason, in section \ref{section3}, we calculate the
contributions from the interaction of four entropic modes mediating
one adiabatic mode to the trispectra. In subsection \ref{3.1}, we
derived the general form of all the 4-point correlation functions,
and in subsection \ref{3.2}, we further analyzed the shape function
and plotted the shape diagrams for two specific momenta
configuration ``equilateral configuration" and ``specialized planar
configuration". And our figures showed that one can easily
distinguish the two types of configurations, because in the
``equilateral configuration" the shape function blows up when
$\bp_1$ was perpendicular to $\bp_{13}$, or equivalent to say when
$p_{13}\rightarrow0$, however, in the ``specialized planar
configuration" it was highly-peaked in the ``squeezed limit"
($p_1,p_2\rightarrow0$).

\begin{acknowledgments}
We thank Xingang Chen, Yi Wang, Eugene A. Lim for many useful
discussions and comments, and KITPC-CAS for the conferences of
connecting fundamental physics with observations. XG is grateful to
Miao Li for reading the manuscript, BH is grateful to Rong-Gen Cai
for the careful reading of the manuscript. BH is supported in part
by the Chinese Academy of Sciences under Grant No. KJCX3-SYW-N2 and
National Natural Science Foundation of China under Grant Nos.
10821504 and 10525060. XG was supported by the NSFC grant
No.10535060/A050207, a NSFC group grant No.10821504 and Ministry of
Science and Technology 973 program under grant No.2007CB815401.
\end{acknowledgments}

\appendix

\end{document}